\documentclass[12pt]{article}
\usepackage{epsfig}
\usepackage{amsmath}
\usepackage{hhline}
\usepackage{amssymb}
\usepackage{times}

\newlength{\dinwidth}
\newlength{\dinmargin}
\begin{document}
\newcommand{\gammaT}{\ensuremath{\gamma^*_T} }
\newcommand{\gammaL}{\ensuremath{\gamma^*_L} }
\newcommand{\xg}{\ensuremath{x_\gamma} }
\newcommand{\xgmy}{\ensuremath{x^{\mathrm{jets}}_\gamma} }
\newcommand{\qq}{\ensuremath{Q^2} }
\newcommand{\GeV}{\ensuremath{\mathrm{GeV}} }
\newcommand{\pb}{\ensuremath{\mathrm{pb}} }
\newcommand{\gevsq}{\ensuremath{\mathrm{GeV}^2} }
\newcommand{\Etone}{\ensuremath{E^*_{T\, 1}} }
\newcommand{\Ettwo}{\ensuremath{E^*_{T\, 2}} }
\newcommand{\Etmy}{\ensuremath{E^*_T} }
\newcommand{\Etsq}{\ensuremath{E_T^2} }
\newcommand{\Etmysq}{\ensuremath{E_T^{*\, 2}} }
\newcommand{\etaone}{\ensuremath{\eta^*_{1}} }
\newcommand{\etatwo}{\ensuremath{\eta^*_{2}} }
\newcommand{\etamy}{\ensuremath{\eta^*} }
\newcommand{\gp}{\ensuremath{\gamma^*}p }
\newcommand{\ycut}{\ensuremath{y_c} }
\newcommand{\PDFgamma}{\ensuremath{D_{i/\gamma^*}} }
\newcommand{\PDFgammaQED}{\ensuremath{D_{i/\gamma^*}^{\mathrm{QED}}} }
\newcommand{\PDFgammaT}{\ensuremath{D_{i/\gamma^*_T}} }
\newcommand{\PDFgammaL}{\ensuremath{D_{i/\gamma^*_L}} }
\newcommand{\PDFproton}{\ensuremath{D_{j/p}} }
\newcommand{\lumi}{57~pb$^{-1}$}
\newcommand{\lambdaQCD}{\ensuremath{\Lambda_{\mathrm{QCD}}} }
\newcommand{\muf}{\ensuremath{\mu_f}}
\newcommand{\mufsq}{\ensuremath{\mu_f^2}}
\newcommand{\mur}{\ensuremath{\mu_r}}
\newcommand{\mursq}{\ensuremath{\mu_r^2}}
\newcommand{\muff}{\ensuremath{\hat{\mu}_f}}
\newcommand{\muffsq}{\ensuremath{\hat{\mu}^2_f}}
\newcommand{\dsqex}{${\rm d}^3\sigma_{\rm {2jet}}/{\rm d}Q^2
{\rm d}\Etmy\, {\rm d}\xgmy$ }
\hyphenation{ca-lo-ri-me-ter}

\begin{titlepage}
\vspace*{2cm}
\begin{center}
\begin{Large}

{\bf QCD Analysis of Dijet Production at Low {\boldmath $Q^2$} at HERA}

\vspace{2cm}

J.\,Ch\'yla, J.\,Cvach, K.\,Sedl\'ak, M.\,Ta\v{s}evsk\'y

\end{Large}
\end{center}

\vspace{2cm}

\begin{abstract}
\noindent
Recent H1 data on triple differential dijet cross sections in
$e^\pm p$ interactions in the region of low photon virtualities
are shown to be in reasonable agreement with the predictions of the NLO
QCD calculations obtained using the program NLOJET++.
The implications of this observation for the phenomenological
relevance of the concept of resolved virtual photon are discussed.
\end{abstract}

\vspace{1.5cm}
\end{titlepage}

\newpage
\section{Introduction}
Recent H1 data on triple differential dijet cross sections in $e^\pm p$
interactions in the region of low photon virtualities \cite{Clanek} have
revealed  a clear excess of the data
over NLO QCD predictions obtained using the DISENT program \cite{DISENT}.
This excess comes predominately from the region of low photon virtualities
$Q^2$ ($Q^2\gtrsim 2$ GeV$^2$) and small $x_{\gamma}$ ($x_{\gamma}\le 0.75$),
where \xg denotes fraction of the four-momentum of the photon carried
by the parton involved in the hard collision
\footnote{ We use the notation and terminology as described in detail
in \cite{Clanek}.}.
We recall that DISENT dispenses with the concept of resolved photon and
evaluates, up the order $\alpha\alpha_s^2$, the direct photon contribution
only. In the region $Q^2\gtrsim 2$ GeV$^2$ this is quite a legitimate
procedure. However, it has been argued \cite{virt,my}
that even for such moderate values of $Q^2$ the concept of resolved
virtual photon is useful phenonomenologically as a way of resuming part
of higher order direct photon contributions. These higher order QCD corrections
are important particularly in the region $x_{\gamma}\le 0.75$, to which only
the lowest order tree level diagrams contribute in DISENT.

\begin{figure}[t]\centering\unitlength=1mm
\begin{picture}(160,50)
\put(20,0){\epsfig{file=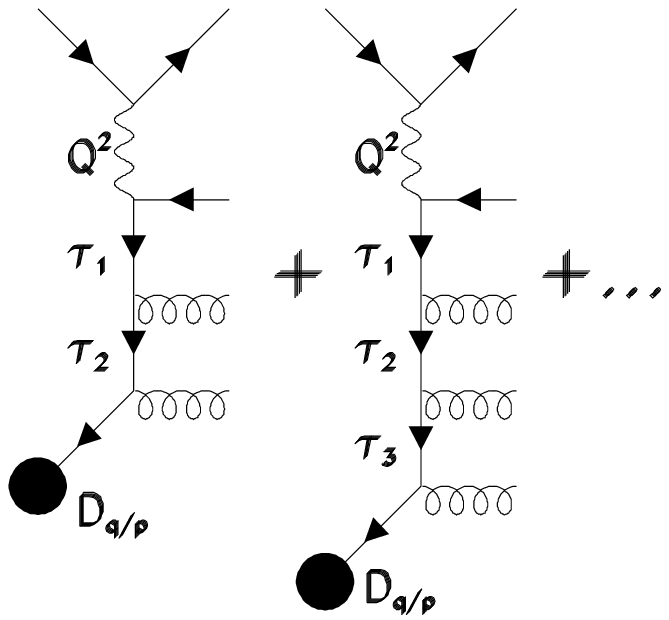,height=5cm}}
\put(80,0){\epsfig{file=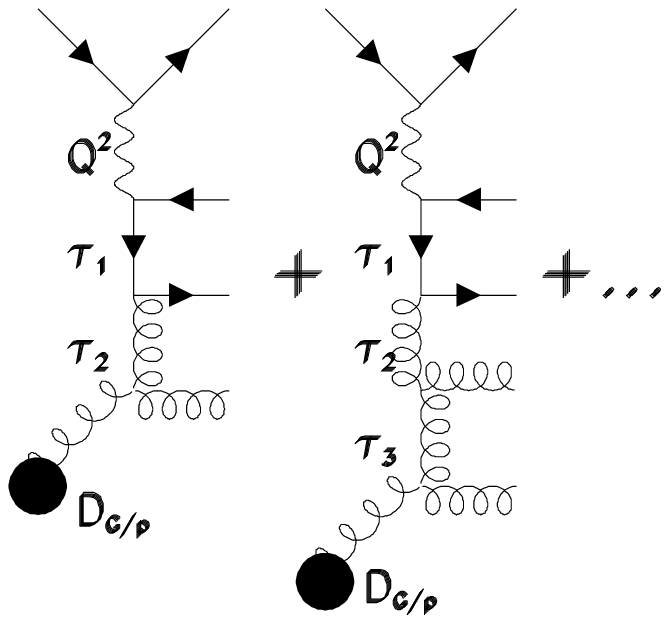,height=5cm}}
\put(35,2){\bf \large a)}
\put(60,2){\bf \large b)}
\put(90,2){\bf \large c)}
\put(120,2){\bf \large d)}
\end{picture}
\caption{Examples of the LO (a) and (c) and NLO (b) and (d) Feynman diagrams
included in NLOJET++ and contributing to the three jet cross section in the
region $x_{\gamma}\le 0.75$. In the framework of the concept of resolved
virtual photon the integrals over the region of small $\tau_1$ give rise to
quark (a-c) and gluon (d) distribution function of the photon.}
\label{res}
\end{figure}

In this region DISENT calculations are described by diagrams
exemplified by those in Fig. \ref{res}a,c. In the kinematic region
$\tau_1\ll \tau_2 \simeq E_T^2$ (where $\tau_1, \tau_2$ denote the
virtualities of the ladder partons in Fig. \ref{res}a,c) the contribution to
the dijet cross section coming from the diagram in Fig. \ref{res}a can be
approximated by the sum of the convolutions which describe the contributions
of transverse ($k=T$) and and longitudinal ($k=L$) polarizations of the virtual
photon.
\begin{equation}
\frac{{\mathrm{d}}\sigma_k(ep\rightarrow jets)}{{\mathrm{d}}y{\mathrm{d}}Q^2
{\mathrm{d}}x_{\gamma}{\mathrm{d}}E_T^2}
\propto f_{\gamma/e}(y,Q^2)\otimes D^{\mathrm{QED}}_{q/{\gamma}^*_k}
(x_{\gamma},Q^2,E_T^2) \otimes \frac{{\mathrm{d}}\sigma^{LO}
(q\overline{q}\rightarrow GG;yx_{\gamma})}{{\mathrm{d}}E_T^2}
\label{res1}
\end{equation}
where
\begin{eqnarray}
f_T(y,Q^2) & = & \frac{\alpha}{2\pi}
\left[\frac{2(1-y)+y^2}{y}\frac{1}{Q^2}-\frac{2m_e^2 y}{Q^4}\right],
\label{tranflux}\\
f_L(y,Q^2) & = & \frac{\alpha}{2\pi}
\left[\frac{2(1-y)}{y}\frac{1}{Q^2}\right].
\label{longflux}
\end{eqnarray}
describe the fluxes of transverse and longitudinal photons in the incoming
electron, the QED contributions to the quark and gluon distribution functions
of the transverse and longitudinal virtual photons are given as
\begin{eqnarray}
D_{q_i/\gammaT}^{\mathrm{QED}}(\xg,Q^2,M^2)&=&\frac{\alpha}{2\pi}3e_i^2
\left(\xg^2+(1-\xg)^2\right)\ln\frac{M^2}{\xg Q^2},
\label{QEDT}\\
D_{q_i/\gammaL}^{\mathrm{QED}}(\xg,Q^2,M^2)&=&\frac{\alpha}{2\pi}3e_i^2
4\xg(1-\xg) \left(1-\frac{\xg Q^2}{M^2}\right),
\label{QEDL}\\
D_{G/\gamma^*_{T,L}}^{\mathrm{QED}}(\xg,Q^2,M^2)&=&0,
\label{QEDg}
\end{eqnarray}
and $\sigma^{LO}(q\overline{q}\rightarrow GG)$ stands for the lowest order
contribution to the cross section of the process $q\overline{q}\rightarrow GG$.
Similar expressions can be written for other partonic subprocesses.
In Eqs.~(\ref{QEDT}-\ref{QEDg}), $e_i$ denotes the electric charge of the
quark $q_i$ and $M$ stands for the factorization scale, in (\ref{res1})
identified with
jet transverse energy. The full expressions for the distribution functions
(\ref{QEDT}-\ref{QEDg}), containing the exact $Q^2$ dependence
with the correct threshold behaviour for $Q^2/m_q^2 \rightarrow 0$, can be
found in~\cite{my}. The contributions of the diagrams in Fig. \ref{res}a,c
contain also singularities coming from the region of small $\tau_2$ and those of
Fig. \ref{res}b,d from the region of small $\tau_3$, but these are understood
to be absorbed in the parton distribution functions of the proton.

\begin{figure}\centering\unitlength=1mm
\begin{picture}(160,30)
\put(30,17){\epsfig{file=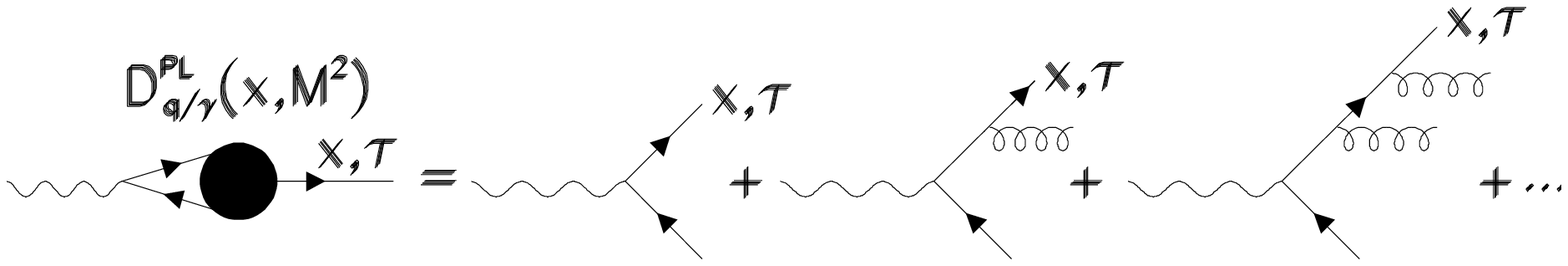,height=1.5cm}}
\put(30,0){\epsfig{file=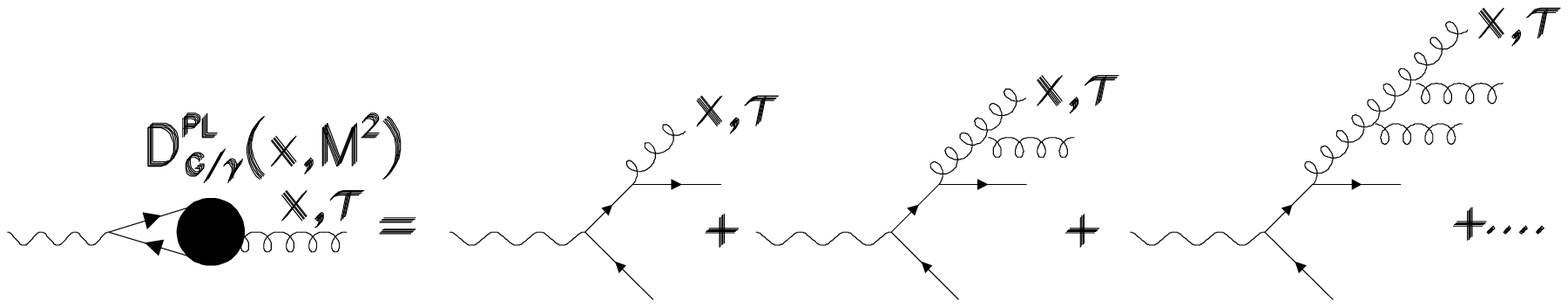,height=1.8cm}}
\end{picture}
\caption{Graphs contributing to the definition of the pointlike part of quark
and gluon distribution functions of the photon.}
\label{pl}
\end{figure}

In \cite{Clanek} it has also been shown that adding in HERWIG
the LO resolved photon contribution to the LO direct one
helps bring the QCD predictions closer to the data. At the NLO such a
procedure can be performed only at the parton level and currently only
within the JETVIP program \cite{jetvip}.
The NLO QCD corrections come loop corrections to LO diagrams like
those of Fig. \ref{res}a,c and from tree Feynman diagrams exemplified by
those of Fig. \ref{res}b,d, which contain terms
proportional to the "large collinear logs" of the form $\ln^2(E_T^2/\tau_2)$.
Within the framework of resolved virtual photon contribution, these terms
can be interpreted as lowest QCD corrections to the purely QED part of quark
and gluon distribution functions of the photon (\ref{QEDT}-\ref{QEDL}).
In Fig. \ref{pl} they correspond to the second (first) term in the definition
of the so called pointlike part of quark (gluon) distribution function
$D_{q/\gamma}^{\mathrm{PL}}$ ($D_{G/\gamma}^{\mathrm{PL}}$) of the photon
\cite{virt}.

Contrary to DISENT, which uses the subtraction method, JETVIP employs the
phase space slicing method to regularize mass singularities. Unfortunately,
the NLO resolved photon calculations obtained with JETVIP turned out
\cite{Kamil} to be
rather sensitive to choice the associated slicing parameter $y_c$.
Consequently, although adding resolved photon contribution worked in the
right direction, no quantitative conclusion could be drawn \cite{Clanek}.

The situation has recently changed due to the appearance of the NLOJET++
program \cite{nlojet}, which allows the user to calculate beside single
and dijet cross sections also the triple jet cross sections to the NLO.
This program uses the same subtraction
method as DISENT and similarly as the latter dispenses with the concept
of resolved virtual photon. As far as single and double jet cross sections
are concerned it is thus in principle identical to DISENT. It, however,
provides also the option -- which we shall call 3-jet mode -- of
calculating NLO QCD correction to three well separated jets. This mode
can be selected by cutting off the region of $x_{\gamma}$
close to 1, where also the dijet final states contribute. Respecting the binning
of the data in \cite{Clanek} we set $x_{\gamma}\le 0.75$.
Comparing the results obtained with NLOJET++ and JETVIP we could address two
questions:
\begin{itemize}
\item how important are terms of the order $\alpha\alpha_s^3$ included in
exact NLO calculations,
\item how important is the all order resummation of the collinear logarithms
into the QCD-improved PDF of the photon used in the resolved
photon contribution of JETVIP?
\end{itemize}
Unfortunately, as mentioned above the NLO resolved photon contribution are
rather sensitive to the variation of the slicing parameter $y_c$.
Moreover, there is also a sizable discrepancy between the results of the NLO
direct photon contribution of JETVIP on one side and those of DISENT or
NLOJET++ in 2-jet mode on the other, with JETVIP results lying in the region
$x_{\gamma}\le 0.75$ significantly below those of DISENT or NLOJET++. This
discrepancy, clearly visible in Fig. \ref{fig.lin}, is not understood.

\section{Data Selection}
The data employed in this article have been published in~\cite{Clanek},
where a detailed description of event selection and experimental methods
used in their extraction can be found. Only the main facts are therefore
recalled. The data, taken at HERA in the years 1999 and 2000, when
electrons with an energy of 27.55~GeV collided with protons with an energy
of 920~GeV, correspond to an integrated luminosity of \lumi.
The kinematic region of the present analysis is defined by cuts on the
photon virtuality $Q^2$ and its inelasticity $y$
\begin{equation}
2< Q^2 < 80~{\GeV}^2,~~~~0.1 < y < 0.85
\label{e1}
\end{equation}
as well as by cuts on the hadronic final state, which has to contain at
least two jets found using the
longitudinally invariant $k_t$ jet algorithm~\cite{INVKT}. The jet transverse
energies, \Etmy\!, and pseudorapidities, \etamy\!, are calculated relative to
the $\gamma^* p$ collision axis in the $\gamma^* p$ center-of-mass frame.
The jets are ordered according to their transverse
energy, with jet 1 being the highest \Etmy jet. The two jets with the highest
transverse energies (leading jets) are required to satisfy
\begin{equation}
\Etone > 7~\GeV,~~~ \Ettwo > 5~\GeV,
\label{e2}
\end{equation}
\begin{equation}
-2.5 < \etaone < 0,~~~ -2.5 < \etatwo < 0.
\label{e3}
\end{equation}
In total 105\,658 events satisfied these selection
criteria. For dijet events the variable
\begin{equation}
\label{rovnice3}
\xgmy = \frac{\sum\limits_{j=1,2}(E^*_{j}-p^*_{z,j})}
{\sum\limits_{\rm hadrons}(E^*-p^*_{z})}\,
     \end{equation}
was used as a hadron level estimate of $x_{\gamma}$. The sum in the
numerator runs over the two leading jets and the sum in the denominator includes
the full hadronic final state.

The data were corrected for initial and final state QED radiation effects, trigger
inefficiencies, limited detector acceptance and resolution and a
photoproduction background. The systematic errors of different origin are added
in quadrature. The dominant source of the systematic error arises from the
uncertainty of the energy calibration of the H1 calorimeters and the model
dependence of acceptance corrections \cite{Clanek}.

\section{Results}
\begin{figure}[b]\centering
\epsfig{file=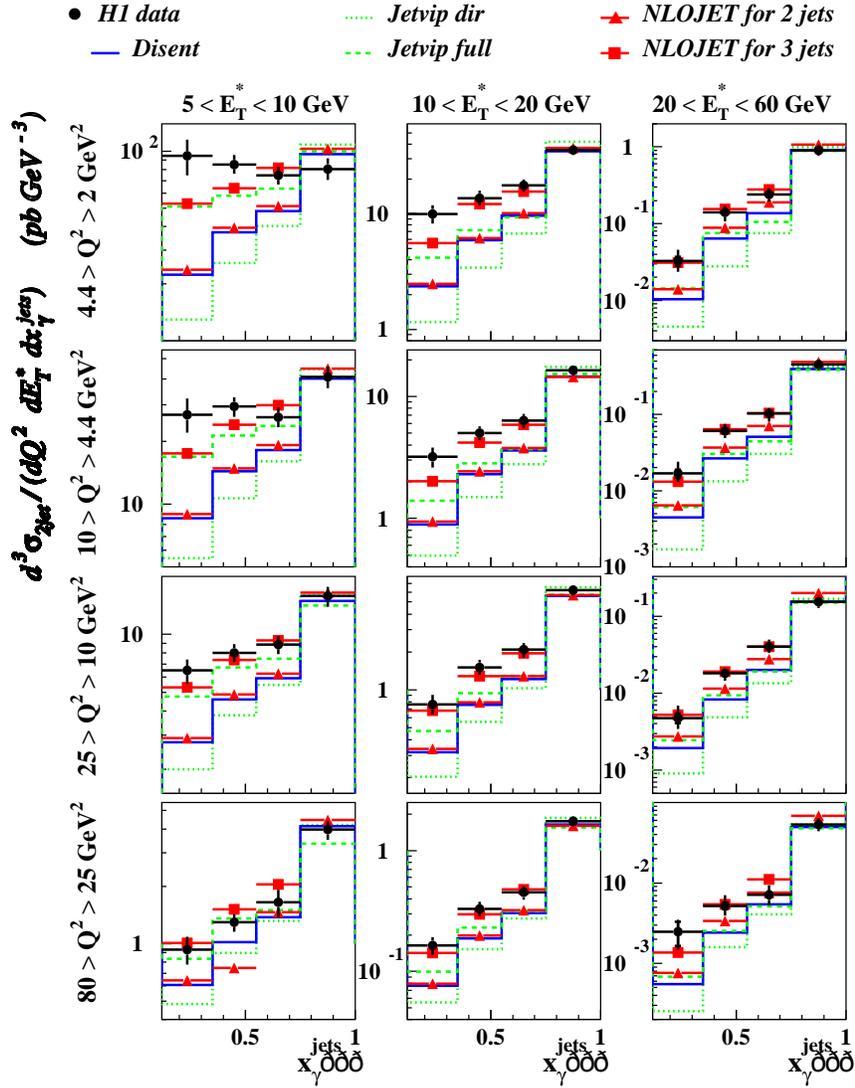,height=14.5cm,%
bbllx=5pt,bblly=15pt,bburx=520pt,bbury=673pt,clip=}
\caption{Triple differential dijet cross section
\dsqex with asymmetric \Etmy cuts (see text). The inner error bars on
the data points show the statistical error, the outer ones
the quadratic sum of systematic and statistical errors. The data are
compared to NLO direct photon calculations using DISENT (full line) and
JETVIP (dotted line), the sum of NLO direct and resolved photon
contributions of JETVIP (dashed line) and the NLOJET++ predictions
in 2-jet (triangles) and 3-jet (squares) mode. All calculations are
corrected for hadronisation effects.}
\label{fig.lin}
\end{figure}

In Fig. \ref{fig.lin} the NLOJET++ results corresponding to both 2-jet and 3-jet
modes and evaluated in the kinematic region (\ref{e1}-\ref{e3}) are compared with
the H1 data \cite{Clanek}. For comparison, the results obtained with DISENT
\cite{Clanek} are plotted as well.

There is a large difference between the NLOJET++ results obtained in 2-jet and
3-jet modes, the latter lying systematically above the former. This
difference, which is most pronounced for small $x_{\gamma}$ and low $Q^2$,
indicates the importance of the NLO QCD corrections in this region.
The NLOJET++ results in the 3-jet mode come significantly closer to
the data than those of the 2-jet mode, but they still undershoot it. The remaining
gap between the data and NLOJET++ calculations in the 3-jet mode is again most
pronounced for small $x_{\gamma}$ and low $Q^2$. In view of large NLO QCD
corrections in this region this remaining discrepancy is not surprising and
and allows room for still higher order QCD corrections.

At the moment the only, though approximate, way of estimating these higher order
terms exploits the concept of resolved virtual photon and the only code that offers
this option at the NLO is JETVIP. Despite the problems mentioned above, we plot in
Fig. \ref{fig.lin} the full JETVIP results \cite{Clanek}, i.e. the sum of direct
(with photon splitting term subtracted) and resolved photon contributions. Ideally,
one would expect the full JETVIP results to be above those of NLOJET++ in 3-jet mode.
Fig. \ref{fig.lin} shows that they are in fact close to each other. This may be
caused by various effects, but we would like to point out the following one. As
already mentioned above the direct photon results of JETVIP are systematically
below those of NLOJET++ in the two jet mode. Assuming the latter are correct
but the resolved photon contribution is estimated correctly by JETVIP, the "correct"
full JETVIP should lie somewhat above the NLOJET++ 3-jet mode predictions and close
to the data. At the moment, this is just a pure speculation, but we hope to check it,
once a new version of JETVIP becomes available \cite{Klasen2}.

In summary, we have shown that the NLOJET++ calculations of the dijet cross sections
in the 3-jet mode are significantly closer to the H1 data \cite{Clanek} that those of
DISENT. This demonstrates the importance of the NLO QCD corrections in processes
involving virtual photons. The remaining gap between the data and current QCD
calculations in the region of low $Q^2$ and small $x_{\gamma}$ may be closed
using the concept of the resolved virtual photon contribution.

\vspace*{-0.5cm}
\section*{Acknowledgements}

We are grateful to Z. Nagy and Z. Tr\'{o}cs\'{a}nyi for extensive correspondence
on some aspects of using their code, to R. P\"{o}schl for help in running
the NLOJET++ code and to Tancredi Carli for drawing our attention to the
3-jet mode of NLOJET++. This work has been supported by the
Ministry of Education of the Czech Republic under the project LN00A006.


\vspace*{-0.5cm}
\begin{thebibliography}{99}
\bibitem{Clanek}
A.~Aktas {\it et al.} [H1 Collaboration],
Eur.\ Phys.\ J.\ C {\bf 37} (2004) 141
[hep-ex/0401010].
\bibitem{DISENT} S. Catani, M. Seymour, Nucl. Phys. {\bf B} 485 (1997), 291
\bibitem{virt}
G.~A.~Schuler and T.~Sj\"{o}strand, Phys.\ Lett.\ B {\bf 376} (1996) 193
hep-ph/9601282,\\
M.~Gl\"{u}ck, E.~Reya and M.~Stratmann, Phys.\ Rev.\ {\bf D} {\bf 54}
(1996) 5515
hep-ph/9605297. \\
M.~Gl\"{u}ck, E.~Reya and I.~Schienbein,
Phys.\ Rev.\ D {\bf 60} (1999) 054019, hep-ph/9903337.
\bibitem{my}
J.~Ch\'{y}la and M.~Ta\v{s}evsk\'{y}, Phys.\ Rev.\ D {\bf 62} (2000) 114025
hep-ph/9912514.
\bibitem{jetvip} B.~P\"{o}tter,
Comput.\ Phys.\ Commun.\  {\bf 119} (1999) 45, {\bf 133} (2000) 105.
\bibitem{Kamil} K. Sedlak, PhD Thesis, Prague, June 2004
\bibitem{nlojet} Z. Nagy, Z. Tr\'{o}cs\'{a}nyi, Phys. Rev. Lett. 87
(2001), 082001
\bibitem{INVKT}
S.~Catani, Y.~L.~Dokshitzer, M.~H.~Seymour, B.~R.~Webber,
Nucl. Phys. B {\bf 406} (1993) 187.
\bibitem{Klasen2} M. Klasen, private communication
\end{thebibliography}
\end{document}